\documentclass[aps,preprint]{revtex4}
\usepackage{amsmath}
\usepackage{amsfonts}
\usepackage{amssymb}
\usepackage{graphicx}
\begin{document}

\title{The excitonic insulator route through  a dynamical phase transition induced by an optical pulse.}
\author{Serguei Brazovskii}
\affiliation{LPTMS, CNRS, Univ. Paris Sud, Universit\'e Paris--Saclay, 91405 Orsay, France}
\email{brazov@lptms.u-psud.fr}
\affiliation{Jozef Stefan Institute, Jamova 39, Ljubljana, SI--1000, Slovenia}
\author{Natasha Kirova}
\affiliation{LPS, CNRS, Univ. Paris Sud, Universit\'e Paris--Saclay, 91405 Orsay, France}
\date{19/12/2015}

\begin{abstract}
We consider a dynamical phase transition induced by a short optical pulse in a system prone to a  thermodynamical instability. We address the case of pumping to excitons whose density contributes directly to the order parameter. To describe both thermodynamic and dynamical effects on equal footing, we adopt for the phase transition a view of the  excitonic insulator  and suggest a formation of the Bose condensate for the pumped excitons. The work is motivated by experiments in donor--acceptor organic compounds with a neutral--ionic phase transition coupled to the spontaneous lattice dimerization and to charge--transfer excitons. The double nature of the ensemble of excitons leads to an intricate time evolution, in particular to macroscopic quantum oscillations from the interference between the Bose condensate of excitons and the ground state of the excitonic insulator. The coupling of excitons and the order parameter  also leads to self--trapping of their wave function, akin to self--focusing in optics. The locally enhanced density of excitons can surpass a critical value to trigger the phase transformation, even if the mean density is below the required threshold. The system is stratified in domains that evolve through dynamical phase transitions and sequences of merging.
The new circumstances in experiments and theory bring to life, once again, some remarkable inventions made by L.V. Keldysh.
\end{abstract}
\maketitle

\section{Introduction}
\subsection{Aftermaths of optical pulses: from Bose--Einstein condensation of excitons to the excitonic insulator.}

Phase transformations induced by short optical pulses is a new mainstream in
studies of cooperative electronic states (see materials of recent meetings \cite{impact,PIPT-meeting,PIPT-proceedings} and the collection \cite{PIPT book}). In experiments on pump--induced phase transitions (PIPT) in electronic systems,  the pumping usually proceeds via transitions among filled and empty electronic bands. A more special and rare technique is the pumping to bound excitations; the excitons whose concentration can reach a very high value of 10\% per unit cell.

In its pure form, such an ensemble of excitons can already show a  number of coherent effects, including their Bose--Einstein condensation (BEC), the idea of which was pioneered by  Keldysh and co--authors \cite{Keldysh:68}. In theory, this prediction was followed and elaborated through decades till nowadays (see, e.g., \cite{Comte:82} and a recent review \cite{Littlewood:04}).
About the same time, the word "exciton" was introduced in another concept, that of an excitonic insulator  \cite{Kohn}, following a vague suggestion in \cite{Knox:63} and its first elaboration \cite{Cloizeaux}.
The excitonic insulator is a hypothetical phase of a semiconductor that appears if the total energy of an exciton $E_{ex}=E_{g}-E_{b}$ vanishes, $E_{ex}\rightarrow 0$. This possibility implies that the conduction gap $E_{g}$ and the binding energy $E_{b}$ can be manipulated (e.g., by pressure or composition) independently.
Soon, it became clear that the excitonic insulator is a mirror part of the  Keldysh--Kopaev state that had already  been suggested\cite{Keldysh:1965}. The "excitonic insulator" became the common nickname for a state formed by the appearance of the electron--hole condensate on top of a semiconducting or  semimetallic state.
A large number of theoretical studies followed soon (e.g., \cite{Koslov,Halperin,Combescot:72,Keldysh:73} in the first wave).
The notion of the excitonic insulator is revived nowadays as a convenient interpretation of phase transitions in various electronic materials \cite{EI-exp:TaNiSe,EI-exp:TiSe2,EI-exp:TmSeTe}. We also recall the old suggestions and attempts to reach the excitonic insulator state by means of extreme conditions such as high magnetic fields (see \cite{Brandt:72,Brandt:76} for experiments and \cite{Brazov:72,Brazov:73} for peculiarities in a theory).

Already in static conditions, the microscopic theory of the thermodynamic excitonic insulator phase just below the transition and the theory of the BEC of optically pumped excitons are closely related, differing mostly by the respective monitoring parameters, either the chemical potential $\mu_{ex}$ or the density $n_{ex}$ of excitons.
For the optical pumping, this duality was strongly emphasized later, around 1990, in a new wave of the theory of intense optical pumping in semiconductors. A more recent publication \cite{Hannewald} offers a good literature review and a systematic refinement of these results. The studies were provoked by observations of the optical Stark effect for a nonresonant pumping (with the photon energy below $E_{ex}$) when the excitonic insulator appears virtually and lasts only in the course of pumping. For what was called a resonant pumping (i.e., above $E_{ex}$ or even above the fundamental edge $E_g>E_{ex}$), the excitonic insulator appeared as a persistent phase \cite{Glutsch}, but the stationary state may not be achievable \cite{Ostreich}. Rather, the system exposes long--lasting large amplitude oscillations, which is in line with the modern knowledge in PIPT.

The arrival of the PIPT science gives a new momentum to studies of ensembles of excitons with opening to coherent effects. By now, experiments were restricted to the so--called neutral--ionic  transitions, but actually the range of realizations is unlimited since all non metallic systems prone to phase transitions have one type of  exciton or another available for pumping.
Recently, we presented \cite{SB+NK:2014,Yi+NK+SB:2015} a phenomenological modeling of spacio--temporal effects expected when optical excitons are coupled to the order parameter of a first--order phase transition, as it happens in the neutral--ionic  case. The phenomenological approach allowed describing  the thermodynamic transition jointly with  the evolution of the optically pumped ensemble of excitons.

In our picture, a quasi--condensate of excitons appears as an inhomogeneous macroscopic quantum state, which then evolves while interacting with other degrees of freedom prone to instability. Via these interactions with soft modes, the excitons are subject to self--trapping (cf. \cite{Rashba} for polarons and \cite{Krivoglaz} for fluctuons), akin to self--focusing in optics. This locally enhances their density, which can surpass a critical value to trigger the phase transformation, even if the mean density is below the required threshold for the global transition. We have recovered dynamical interplays of fields such as the collective wave function of excitons, the electronic charge transfer and polarization, and the lattice dimerization. We have found various transient effects: self--trapping, dynamic formation of domains separated by walls, subsequent merging of domains and collapse of walls, and emittance of propagating wave fronts.

That model and the results could be applicable to situations where the excitons and the order parameter are essentially different while interacting  fields.
This could be the case of pumping to high--energy intramolecular excitons in donor--acceptor systems with the neutral--ionic  transition, as it has been realized experimentally in \cite{Koshihara}.

In this article, we consider the case where the transition order parameter and the intensity of pumping excitations are of the same origin as it happens for the low--energy charge--transfer excitons \cite{Okamoto} in the neutral--ionic  transition.   The BEC of excitons is involved in both situations, but the last case also brings to  light  the excitonic insulator state coupled to the BEC.

To describe both thermodynamic and dynamical effects on the same root we adopt a view of the excitonic insulator for the phase transition.
With only the main ingredient, the vanishing of the excitation energy, the concept is too broad, as just a generic view of quantum phase transitions in electronic systems. The focused concept of the excitonic insulator is distinguished when the number of excitons, both in the ground state and out--of--equilibrium, is approximately conserved. (If it is conserved precisely as  had been stated in most theories before the clarifying work by  Keldysh and co--authors \cite{Keldysh:73}, then the thermodynamics of the phase transition is not affected essentially but there would be no dynamical path to the excitonic insulator state, which is of particular importance in the context of PIPT; see precisions in the next section.)

The theory of PIPTs faces great challenges when started \textit{ab initio} at the microscopic level (see, e.g., \cite{Yonemitsu:2012} for a review and \cite{Ishihara}). But over longer time scales, the evolution should be governed by collective variables like the order parameter and lattice deformations. The effectiveness of such a phenomenological approach has been proved by a detailed modeling of coherent dynamics of a macroscopic electronic order through destruction and recovering of the charge density wave state. That allowed  describing  effects such as dynamic symmetry breaking, stratification in domains and subsequent collapses of their walls, all in detailed accordance with  experiment (see e.g. \cite{Yusupov:2010}). Another example was the modeling \cite{Brazovskii:2014-JSNM} of the recently discovered \cite{Stojchevska:2014} switching to a truly stable hidden state of a polaronic Mott insulator in $1T--TaS_2$. The phenomenological approach becomes inevitable when considering spacially inhomogeneous regimes that ultimately appear here. That is what we  keep using in the presented study.

\subsection{The neutral--ionic transition as an excitonic insulator.}\label{ss:NI+EI}
Relevant neutral--ionic  transitions occur in bi--molecular donor--acceptor chains (D$^{-\rho}$--A$^{\rho}$, in particular  TTF--CA, see the references in \cite{Okamoto},\cite{Kobayashi}) that show a variable charge transfer $\rho$ between the lower  $\rho_N$ in the quasi--neutral  high temperature  phase and a higher $\rho_I$ in the low--temperature ionic  phase.
The first--order transition in $\rho$  alone would go without a symmetry breaking and could be described by a generic double--well curve for the free energy $W(\rho)$ with two minima at neutral and ionic states.  In spite of the essential distance  ($\rho_N$=0.32 and  $\rho_I$=0.52 at the phase coexistence), the separating barrier is small, and hence we deal with a first--order phase transition that is close to a second--order one.
That is confirmed by observations of a critical increase in the dielectric constant \cite{Kobayashi}  as a precursor of ferroelectricity and of the Kohn anomaly (see \cite{DAvino} and the references  therein) as a precursor for the lattice dimerization instability.
More richness comes from another degree of freedom, the  alternating molecular displacements $h$: the ionic phase is accompanied by  lattice dimerization, and hence there is a symmetry breaking and the transition could have been of the second order, which is not the case nevertheless: the jump in $h$ is concomitant with the jump in  $\rho$.

Remarkably, TTF--CA and related materials possess two types of observable and treatable excitons: the intramolecular   Frenkel--type excitons as a high--energy (2.33eV) mode of the TTF molecule  and the low--energy (0.6 eV) charge--transfer excitons. In the last case, an electron is activated from the predominantly donor--formed band to the acceptor--formed band,  but the electron and the hole are kept bound as for Wannier--Mott excitons in semiconductors. The charge--transfer exciton  increases the charge disproportion $\rho$ above its initial value, coming from the simple hybridization of donor--  and acceptor-- originated bands. Thus, the neutral--ionic  transition can be viewed as accumulation of virtual e--h pairs in the ionic ground state. With the charge--transfer exciton as the bound state lying well below the unbound e--h threshold $E_g\approx 1.5 \ eV$, the charge transfer can be seen as coming predominantly from excitons, whence the picture of the excitonic inuslator.
The cases of intramolecular exciton  and charge--transfer   exciton   correspond to profound experimental studies by Koshihara \cite{Koshihara} and  Okamoto \cite{Okamoto} with coauthors. Our earlier theoretical work \cite{SB+NK:2014,Yi+NK+SB:2015} can refer to the case of the intramolecular exciton.

We underline the quantum nature of the charge--transfer  exciton in both its internal structure and  motion. Internally, this is a symmetric superposition of states where dipole dimers are formed by the charge transfer from a donor site to the surrounding acceptor sites. Within the classical illustration, that is used sometimes, there would be either left-- or right-- directed dimer with corresponding opposite electric polarizations. But with the quantum superposition, the dipole moment vanishes, to be weakly restored after  lattice dimerization, which breaks the inversion symmetry and mixes states of excitons of even and odd parities. With respect to the motion as a whole, the lowest--energy state of the exciton is ideally (up to inhomogeneities as we  see below) a plane wave delocalized over the sample in the state with the momentum $k=0$, like in condensates of polaritons (see \cite{Berloff:2013} for a review). This description is complementary to the popular classical picture of classical domain walls (the solitons) propagating as a falling domino array.

\section{The model}
\subsection{Integration of the Bose condensation of excitons and of the excitonic insulator in dynamics of neutral--ionic transitions}
In the case of pumping into the intramolecular exciton  near the neutral--ionic  transition, the excitons and the order parameter were essentially different while interacting fields, and it was therefore quite straightforward to formulate a phenomenological model \cite{SB+NK:2014,Yi+NK+SB:2015}. In case of the charge--transfer   exciton, both the excitation and the long--range ordering are built from processes of electronic transfer between donor and acceptor molecules. Although the number of independent fields is reduced, conceptually this case is more intricate.

The  dualism of the exciton density $q$ and  the thermodynamic order parameter $\rho-\rho_N$ already outlined in Section \ref{ss:NI+EI} does not seem  to be quite compatible: the thermodynamic charge transfer  is given by a redistribution of the charge density $\rho$, which is the single real field both in equilibrium and in evolution. The charge--transfer     density from the pumped ensemble of excitons is given by the exciton number density $q\propto |\Psi|^{2}$, and hence the field is still real but its evolution is given by the complex wave function $\Psi=q^{1/2}\exp(i\varphi)$ of the BEC, and a hidden degree of freedom --- the phase $\varphi$ --- comes to the sight. The evolution of this phase can be traced directly because its time derivative gives the observable instantaneous energy of the exciton $E_{ex}(t)=-\hbar\partial_{t}\varphi$ in the frame of their ensemble.

This dualism between the density of microscopic excitons and the thermodynamic charge ordering calls for refining another dualism: among explicit coherent oscillations of the order parameter and those of the wave function of excitons, which interfere but keep the different origins.
In general, collective oscillations might be superimposed on the frequency $E_{ex}(t)/\hbar$, but now, with strong variations of $E_{ex}$, the two time dependences cannot be disentangled and must be considered on equal footing.

\subsection{The energy functional and evolution equations.}
To build a unified approach to two faces of the charge transfer, we describe the phase transition as the one of  the excitonic insulator  state --- a view that is becoming popular, nowadays, as one can see from the papers we have cited above. The notion of the excitonic insulator can be applied to a large category of quantum phase transitions where the instability comes from the vanishing of the energy $E_{ex}$ of an excitation which is here a bound state of the $e-h$ pair. The instability for negative $E_{ex}<0$ is compensated by repulsion of excitons which determines the equilibrium concentration. A particular convenience is that theories of the excitonic insulator and BEC  are identical except that the first is monitored by the chemical potential while the second one  by the mean density. The time evolution generalizes and unifies both views, which we  exploit in what follows.

The increase in  the average charge--transfer intensity does not break the symmetry, similarly to the liquid--vapor transition; a first--order phase transition is then expected in general. (In systems of our interest, there is also a discrete symmetry breaking thanks to appearing of the lattice deformations  $h$, which can take the values $\pm h_0(\rho)$ in equilibrium at a given $\rho$; but for the sake of transparency we disregard this variable for a while.) The energy functional $W(\rho)$ is minimal in equilibrium, which can occur at two values $\rho_{1}$ and $\rho_{2}$, one of which is metastable except  at the transition temperature $T_{c}$ of the first--order transition, where $W(\rho_{1})=W(\rho_{2})$.
At short times of PIPTs and/or at low temperature, a system described by $\rho$ alone would behave dynamically as is described by a Hamiltonian containing the kinetic energy density proportional to $\dot{\rho}^{2}$. It leads to a second--order differential equation $\partial_{t}^{2}\rho\propto-\delta W/\delta\rho$, which does not preserve $\rho$ at all and does not result in a  bottleneck for transformations among phases with different mean values of $\rho$. In conditions of PIPT, such a system performs large--amplitude pendulum oscillations (see a clear experimental example in \cite{Yusupov:2010}). With some dissipation taken into account, it is eventually driven  towards one of equilibrium states $\rho_{1}$ or $\rho_{2}$.

Actually, the optical pumping gives rise initially to a high density of excitons, which, in case of  resonant pumping or after  relaxation in general, can be described by the common wave function $\Psi$ of the quasi--condensate with the density $q=\left\vert\Psi\right\vert ^{2}$ contributing to $\rho=\rho_{1}+q$. The system of excitons itself can be described as interacting bosons whose phenomenological treatment at low temperature can be based upon the adapted Gross--Pitaevskii  theory. And that would lead to a kind of  nonlinear Schr\"odinger equation (NLSE), which commonly preserves the number of excitons, and  would therefore prohibit any evolution of $\rho$.

The escape from these contradictions can be found following the work by Keldysh and co--authors  \cite{Keldysh:73}: there are processes of creation and annihilation of pairs of excitons from/to the vacuum coming from matrix elements of the Coulomb interaction, which transfer two electrons across the gap, between filled and empty bands, (see  Fig.1).
\begin{figure}[hptb]
\includegraphics[width=5cm]{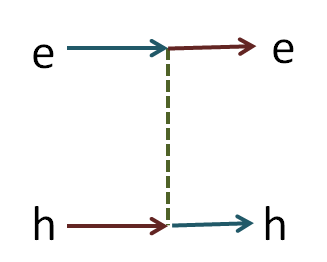} \quad
\includegraphics[width=5cm]{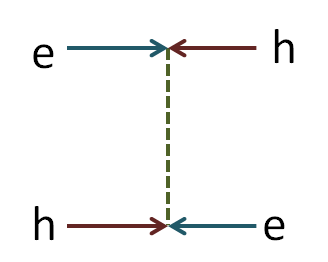}
\caption{Coulomb interactions of the electron and the hole: normal (left panel) and anomalous, with the annihilation of excitons' pairs (right panel).}
\label{fig1}
\end{figure}

That gives rise to the amplitude $S$ of  simultaneous annihilation of two excitons. Finally, the  free energy of the excitonic insulator acquires the phase--fixing terms
\[(S^{\ast}\Psi^{2}+S\Psi^{\ast2})/2\ ,\ S=|S|e^{(i\alpha)}\]
and hence the generalized Ginzburg--Pitaevskii equation, (see below) does not preserve the total number of particles. (Here and hereafter, we  assume the given phase $\alpha=0$ which can always  be done by shifting the origin of the variable phase $\varphi$.)
For the founding excitonic insulator scenario of condensation of Wannier--Mott type excitons, these anomalous  terms are  small compared with the dominant Coulomb energy $E_{b}$ as   $|S|/E_{b}\propto(a/R)^{d}$ where $a$ is the lattice spacing, $R\gg a$ is a large exciton radius, and $d$ is the space dimension. For local excitons with $R\propto a$, there is no smallness; the phase can be strongly fixed and the system must show the behavior expected of the generic scalar order parameter.

By the definition of the excitonic insulator, the phase--fixing terms are small, $|S|\ll 1$, and hence the total complex order parameter $\Psi$  still is to  be exploited. For a typical neutral--ionic  material, the bare value of $S$ may not be very be small: both because $R$ is only of a few $a$ and because of low $d=1$. But the same fact that the system is nearly one--dimensional brings, as always in one dimension, strong phase fluctuations that reduce functions periodic in $\varphi$. Hence, the effective value
$|S|\rightarrow|S| \langle exp(2i\varphi) \rangle \approx|S|\exp(-2\langle\varphi^{2} \rangle)$
can be small; it can even be renormalized to zero.

We  work with a special form of the Ginzburg--Pitaevskii equation that is applicable when boson occupation numbers for all relevant states are much bigger than unity. For the system of excitons on a d--dimensional lattice, the condition is that their mean density per lattice site is $x\gg (T/D)^d$. With the exciton bandwidth $D\sim 10^3 \ K$, this inequality can  always be satisfied for a typical experimental value $T\sim 10^1$. Even if the initial value of $x$ is not sufficiently high, the kinetics of many--particle cooling feeds the low--energy states such that the Ginzburg--Pitaevskii theory becomes applicable sooner or later. This is an advantage of the fast PIPT technique, where the integral time of observations is shorter than the recombination time of excitons.

Varying  the energy functional $W$ (to be specified in Eq. (\ref{V(q,h)}) below) over $\Psi$ yields the generalized Ginzburg--Pitaevskii    equation

\begin{eqnarray}
i\hbar\partial_{t}\Psi+i\hbar\Gamma\Psi=
\frac{\delta H}{\delta\Psi^{\ast}}= -\frac{\hbar^{2}}{2M}\partial_{x}^{2}\Psi+V(q)\Psi -S\Psi^{\ast}, \\
V(q)={dW}/{dq}=E_{ex}
\nonumber
\label{psi}
\end{eqnarray}

The perturbations proportional to $\Gamma$ (see below Eq. (2)) and $S$   describe the respective relaxation of  the amplitude and  locking of the phase. The relaxation rate $\Gamma$ might have a complicated behavior, passing through different regimes. In the most dilute limit of isolated excitons, apparently $\Gamma=1/\tau_{ex}$ is the inverse life time for the exciton recombination. Actually, our vanishing $q$ still assumes a macroscopic concentration when the radiational recombination is dominated by  stimulated emission; then, according to the Bose--Einstein statistics, $\Gamma$ decreases as $\Gamma\propto q$. Approaching the high--$q$ equilibrium phase at $q\approx q_{0}$, where the excitons constitute the ground state, $\Gamma\rightarrow 0$ should vanish since  there is no channel of decay at the energy minimum. That is rigorously true below the neutral--ionic  transition when the high $q$ state has a minimal energy. If it is metastable, then we neglect its evaporation over the barrier towards the $q=0$ region. If there were no phase dependence through the $S$ term, then simply $\Gamma$ should vanish together with $V$, and hence we can write a qualitative interpolation between the two limits as $\Gamma\Rightarrow G(q)qV(q)/\hbar$, where $G(q)$ is some structureless dimensionless function of $q$ (which we  take as a constant in the illustrative numerical modeling). This expression tells us that $\Gamma<0$ (meaning  amplification instead of the decay) in the region between the barrier and the high--$q$ minimum where $V<0$, (see Fig.2). 

\begin{figure}[hptb]
\includegraphics[width=0.3\textwidth,height=4cm]{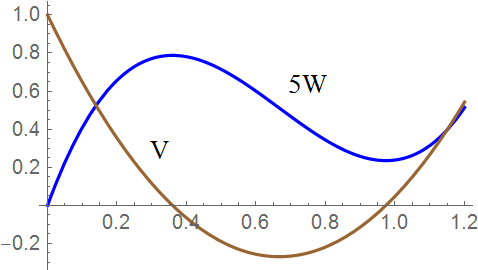}\quad
\includegraphics[width=0.3\textwidth,height=4cm]{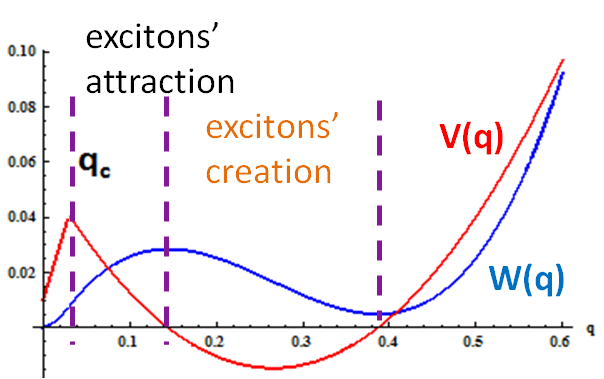}\quad
\includegraphics[width=0.3\textwidth,height=4cm]{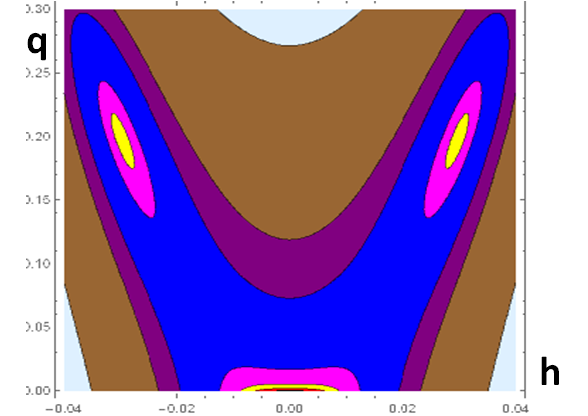}
\caption{Plots for the ground state energy $W(q)$ and the potential $V(q)=dW/dq$ above the first--order transition. Left panel: for a metastable generic excitonic insulator.
Middle panel: for the neutral--ionic system after minimization of $W(q,h)$ over lattice displacement $h$. Vertical dashed lines separate four intervals of q (from the left to the right): repulsion, attraction,  creation, and again repulsion of excitons.
Right panel: density plot of $W(q,h)$ showing all three locally stable states.}
\label{fig2}
\end{figure}

This is not unphysical since here  the system is indeed unstable with respect to spontaneous creation of excitons whose energy becomes negative as  in the excitonic insulator. Bringing the system to this range of $q$ by  pumping  the excitons is similar to instantaneous crossing the boundary of the excitonic insulator state by  varying a thermodynamic monitoring parameter.

In view of the phase dependence, the equilibrium state is determined by both $q$ and $\varphi$ approaching the energy minimum $\varphi=0$ (modulo $\pi$), $q\approx q_{0}$ in some complicated way. Instead of guessing $\Gamma$ as a function of two variables, it is more instructive and basic to realize that the energy relaxation terminates when $\partial_{t}\varphi=0$. Then
\begin{equation}
\Gamma\Rightarrow
-\frac{G}{2i}(\Psi^{\ast}\partial_{t}\Psi-\Psi\partial_{t}\Psi^{\ast})
=-Gq\partial_{t}\varphi=\frac{Gq}{\hbar}E_{ex}(t,x),\ G\approx const.
\label{Gamma}
\end{equation}
The equations for $q$ and $\varphi$ written below show that this expression is indeed a generalization of the relation $\Gamma\Rightarrow G(q)qV(q)$.

For the zero dimension $D=0$ of a quantum dot or for a spacially homogeneous regime $\partial_{x}\Psi\equiv0$, it is instructive to write the above equations in the variables $q,\varphi$ ($\partial_{t}\varphi=\dot{\varphi}~,~\partial_{t}q=\dot{q}~$):

\begin{eqnarray}
\hbar\dot{\varphi}=-V+|S|\cos(2\varphi) \\
\dot{q}=-\Gamma q+2\left\vert S\right\vert q\sin(2\varphi)
=Gq^{2}\dot{\varphi}+\left\vert S\right\vert q\sin(2\varphi)\\ 
\nonumber
=-(G/\hbar)q^{2}(V-|S|\cos(2\varphi))+2\left\vert S\right\vert
q\sin(2\varphi).
\label{d=0}
\end{eqnarray}

The polar trajectory $q(\varphi)$ is given by the solution of the equation

\begin{equation}
\frac{dq}{d\varphi}=
\frac{q^{2}(V-|S|\cos(2\varphi))G/\hbar-\left\vert
S\right\vert q\sin(2\varphi)~}{V-|S|\cos(2\varphi)}
\end{equation}

Approaching the neutral phase with a residual but still macroscopic density of excitons, $q\rightarrow0$, $V\rightarrow E_{ex}^0$, we obtain the vanishing concentration of isolated excitons with the shifted energy $E_{S}=\sqrt{E_{ex}^{02}-|S|^2}$. The wave function oscillates as
\begin{equation*}
\Psi\propto\left(\sqrt{E_{ex}^0-|S|}\cos(\omega_{S}t)+
i\sqrt{E_{ex}^0+|S|}\sin(\omega_{S}t \right),
\end{equation*}
 where $\omega_{S}=E_{S}/\hbar$.
For the high--$q$ phase, unlike the $q=0$ one, the energy has a smooth minimum at $q_0$ where $V(q_0)=0$; then $\partial_{t}\Psi=0$ implies that

\begin{equation*}
\sin(2\varphi)=0, \  \ -V+|S|\cos(2\varphi)=0,
\end{equation*}
i.e.
\begin{equation*}
\varphi_{0}=\varphi=\pi n/2  \ ,\ V(q_{0})=|S|(-1)^n
\end{equation*}
We note that the equilibrium value  is displaced from $q_0$ by the effect of the $S$ term.

\subsection{Taking account of dimerizations}
We now  come to the realities of neutral--ionic  transitions by recalling the symmetry breaking order parameter, the dimerization $h$. We work with the energy function
\begin{eqnarray}
W(q,h) &= E_{ex}^0 q+\frac{a}{2}q^{2}+\frac{b}{3}q^{3}+\frac{d}{2}
(q_{h}-q)h^{2}+\frac{f}{4}h^{4},  \notag \\
V(q,h) & =\frac{\partial W(q,h)}{\partial q} = E_{ex}^0+aq+bq^{2}-\frac{h^{2}d}{2}
\label{V(q,h)}
\end{eqnarray}
The phenomenological model of the generic excitonic insulator would
contain only the terms without the symmetry breaking field of displacements $h$;
then, to obtain the regime with the (meta) stable state at $q_{0}>0$ coexisting with
 the still stable state at $q=0$, we might assume the negative 
 $a<a_{cr}=-2\sqrt{bE_{ex}^0}<0$ (attraction of excitons), but  the  positive $E_{ex}>0$.
In applications to neutral--ionic  transitions, we leave $a>a_{cr}$
(it can even  be  positive $a>0$, corresponding to the repulsion of
excitons, which guarantees  local microscopic stability), because the energy minimum
at $q>0$ is built with the help of the induced instability, at all $q>q_{h}$, of the
field $h$ describing the dimerization. The effect of $h$ cannot depend on its sign,
whence the coupling proportional to $-qh^{2}$; it can be viewed as a decrease in the exciton
energy by  dimerization, which can come from the mixing of even and odd excitonic
states when the inversion symmetry is broken at $h\neq0$. As a second--order
perturbation in $h$, it should be negative as we have specified. We can also add  a higher order coupling proportional to $-q^{2}h^{2}$ which is the effect of dimerization on the interaction of excitons. In the presence of $h\neq0$, the inversion symmetry is broken (the state is ferroelectric!), and the excitons acquire dipole moments along the chain, whence the attractive dipole--dipole contribution, which guarantees the negative sign of this interaction. Taking both effects into account, we can write the interaction term as $(q_{h}-q)(d_{1}+d_{2}q)h^{2}/2$, with $d_{1},d_{2}>0$.

The equation for $h(t,x)$ follows from the variation over $h$ of the energy functional (\ref{V(q,h)}) augmented by the lattice kinetic energy and elasticity:

\begin{equation}
K(\partial_{t}^{2}+
\gamma\partial_{t})h-Ks^{2}\partial_{x}^{2}h+d(q_{}-q)h+fh^{3}=0
\label{h}
\end{equation}
Here, $s$ is the sound velocity and $q_{h}d/K=\omega^{2}$, where $\omega$ is the $h$--mode frequency in the virgin state $q=0$. Equations (\ref{V(q,h)}, (\ref{h}), and (\ref{psi}) with $V(q)$ generalized to $V(q,h)$ from (\ref{V(q,h)}) constitute the full system used in our minimalistic modeling.

\section{Results of  numerical modeling}
\subsection{Generic excitonic insulator}

We first  consider  the generic model of the excitonic insulator schematically. Here, the transition is of the second order, governed by only one field $\Psi$, and the phenomenological energy has the simplest form $W(q)=E_{ex}^0q+aq^2/2$, $V(q)=E_{ex}^0+aq$.
In the excitonic insulator phase, $E_{ex}^0<0$. We select the equilibrium position to be $q_0=-E_{ex}/a=1$ and hide the coefficient $a$ in time rescaling. In the thus reduced equations, we choose $S=0.1$ and fix the attenuation coefficient in (\ref{Gamma}) as $G=0.01$. The pumping intensity determines the initial value $q_i$.
The results are shown in Fig.3 as linear plots for $q(t)$,  $\varphi(t)$ and $E_{ex}^0 (t)$ and parametric polar plots for $q(\varphi)$.

\begin{figure}[htbp]
\includegraphics[width=5cm,height=4cm]{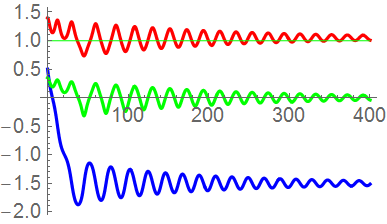} \quad
\includegraphics[width=5cm,height=4cm]{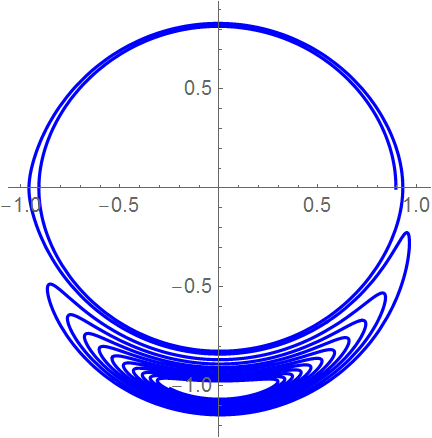}
\caption{Plots for $t$ dependencies (left panel: $q$ in red, $\varphi$ in blue, $E_{ex}^0 (t)$ in green) and the polar trajectory (right panel: $q(\varphi)$) for the generic excitonic insulator after an additional pumping.}
\label{fig3}
\end{figure}

There is a small critical deviation $q_i-q_0$ (to any side) beyond which the phase is unlocked (the $S$ term is not effective), $q$ oscillates with little attenuation around $q_i$, and  the phase rotates almost linearly in $t$ (superimposed by oscillations as well as $q(t)$). This regime corresponds to the collective mode of excitons with oscillations coming from a \emph{macroscopic quantum interference} due to the particle production to/from the excitonic insulator ground state. But with time the attenuation proceeds towards the energy minimum, $q$ crosses the critical deviation towards $q_0$, and the phase locks via a dynamic transition.  With attenuating oscillations,the phase approaches an equilibrium value from the sequence $\pi(n+1/2)$, where $n$ is the number of half--periods processed before the locking.  Also with attenuated oscillations, $q(t)$ approaches the equilibrium value displaced from $1$ by effect of the $S$ term.

Figure 3 shows the time dependences for the phase and the amplitude, and the trajectory as a parametric polar plot for $q(\varphi))$. The initial deviation (pumping from the equilibrium $q_{eq}=1.05$ to some initial $q_i=1.4$ at a given equilibrium phase $\varphi=\pi/2$) provokes the unlocked regime, which lasts until $t\approx 40$ with nearly two (there can be many) rotations over the initial circular trajectory. Here, the amplitude is close to a constant (the number of excitons in the condensate is nearly conserved), while the phase  decreases almost linearly in time, with a nearly constant exciton energy. With $q(t)$ slowly decreasing because of relaxation in the number of excitons,  a locking transition takes place after which the amplitude gradually returns to  thermodynamic equilibrium while the phase is locked at a new allowed value $\pi(n+1/2)$, here with $n=-3$.

\subsection{Generic first--order phase transition}

We now turn to the case of a generic first--order phase transition which may be due to attraction of excitons, as is sometimes considered for semiconductors \cite{Keldysh:68,Keldvsh:88,Nozieres:82}. This example also builds a bridge to our primary goal of a multi--field system. We choose the ground state energy as
 \[W(q)=q((q - 1)^2 + 0.05)/1.05\ , \ V(q)=dW/dq\]
which is plotted in Fig.2.
$W(q)$ is normalized such that the bare exciton energy is $E_{ex}^0=V(0)=1$.
With these parameters, we are below the thermodynamic phase transition to the excitonic insulator state, which nevertheless can exist as a metastable state (the minimum of $W(q)$ at $q_0\approx 1$).

For a very high pumping exceeding the position of the metastable excitonic insulator, $q_i>q_{0}$, the behavior shown in Fig.4 is qualitatively similar to the  above case of the generic excitonic insulator with the second--order transition, Fig.3.

\begin{figure}[htbp]
\includegraphics[width=5cm,height=5cm]{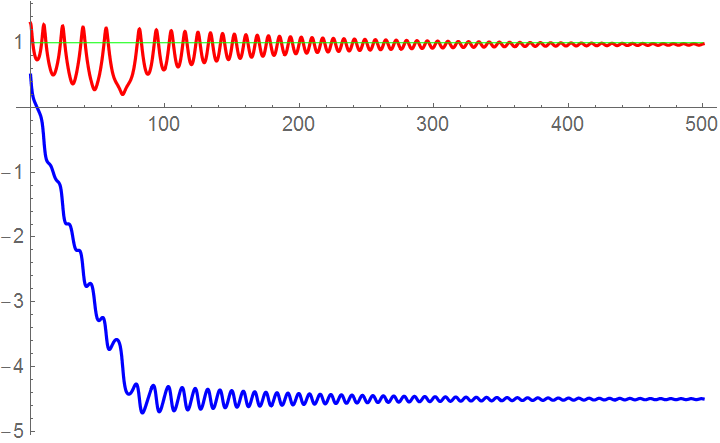} \quad
\includegraphics[width=5cm,height=5cm]{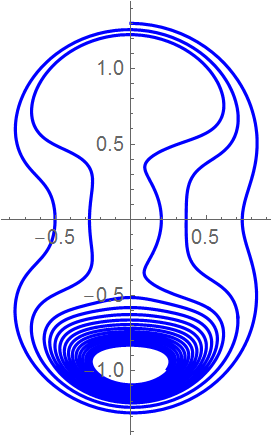}
\caption{Plots for $t$ dependencies (left panel) and the polar trajectory (right panel) for the  metastable excitonic insulator after a high pumping to $q_i=1.3>q_0$.}
\label{fig4}
\end{figure}

At a lower pumping, but still above the position of the barrier in $W(q)$, 
$q_0>q_i>q_b\approx 0.36$, there is a fast crossover to the regime of oscillating relaxation towards the excitonic insulator state, with no clear unlocked regime, as demonstrated in Fig.5 for $q_i=0.35935$ (just above the barrier). 

\begin{figure}[htbp]
\includegraphics[width=0.3\textwidth,height=4cm]{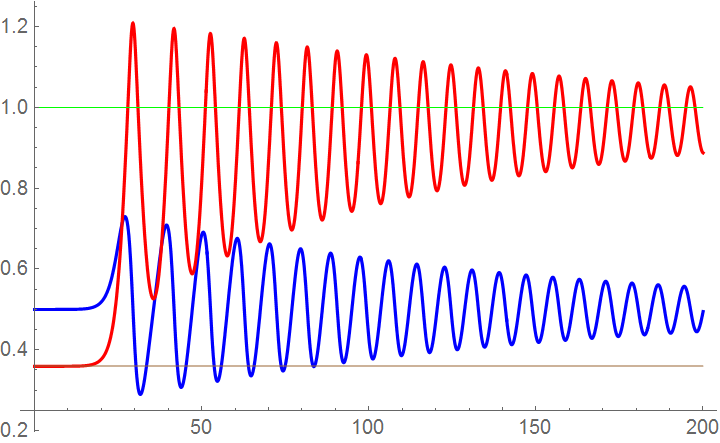} \quad
\includegraphics[width=0.3\textwidth,height=4cm]{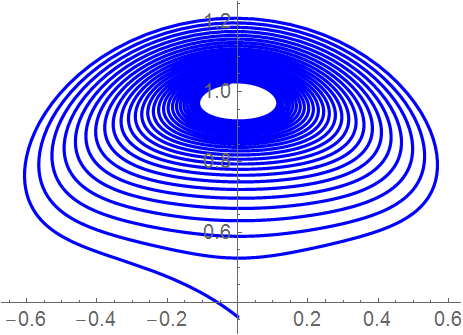} \quad
\includegraphics[width=0.3\textwidth,height=4cm]{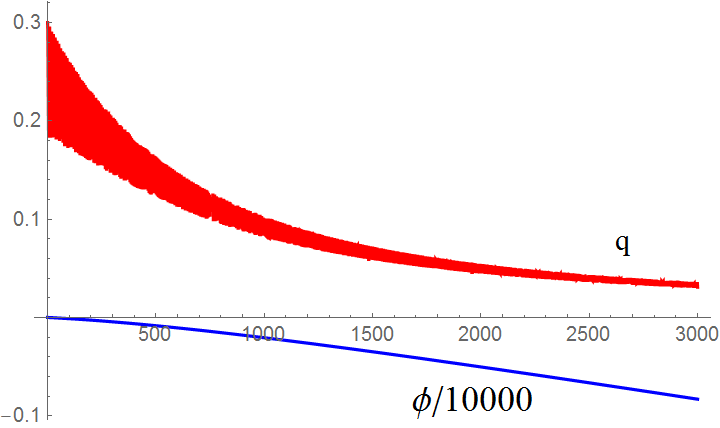}
\caption{Plots for $t$ dependencies (left panel) and the polar trajectory (middle panel) for the  metastable excitonic insulator after a just super--barrier pumping: $q_i=0.35935$. The right panel shows dependencies of $q(t)$ and $\varphi(t)$ for the sub--barrier pumping, $q_i<q_b$.}
\label{fig5}
\end{figure}

After some waiting time (which is pronounced here because of the chosen close proximity to the critical pumping), strong oscillations develop in both $q$ and $\varphi$. After a long relaxation accompanied by attenuating oscillations, $q$ finds a new equilibrium at the position of the metastable excitonic insulator while the phase returns to the initial value $\pi/2$.
For an even lower sub--barrier pumping $q_i<q_b$, the system relaxes to the virgin no--exciton state, as is also shown  in Fig.5. The curves are superimposed by oscillations, which, at least at sufficiently long $t$, correspond to the energy $E_{ex}^0$ of the bare exciton.

\subsection{Multi--field model for the neutral--ionic transition.}

The modeling is based on Eqs. (\ref{psi}), (\ref{V(q,h)}) and (\ref{h}) describing a coupled evolution of the complex order parameter $\Psi$ and of the dimerization $h$. Numerically, we try to stay within the range of parameters known or estimated for the real material with the neutral--ionic  transition. The plausible numbers are given in the Appendix.

\subsubsection{Neutral--ionic system at space independent conditions.}

First, we generalize the generic models considered previously by looking for a space--independent solution. The time dependences are shown in the plots in Fig.6; they have been calculated for the realistic parameters described in the Appendix. 

\begin{figure}[hptb]
\includegraphics[width=0.3\textwidth,height=5cm]{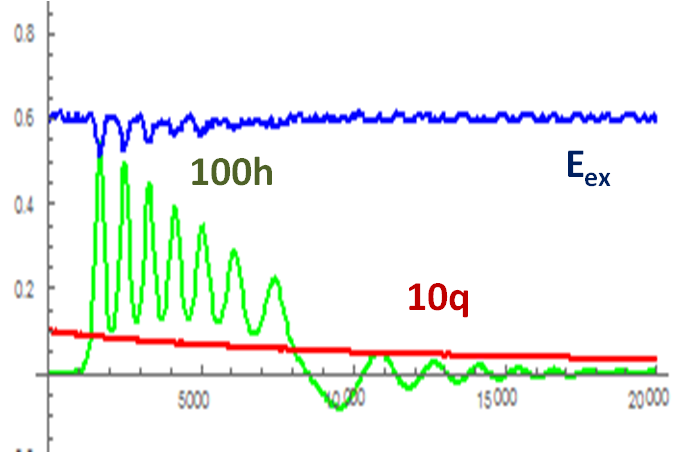} \quad
\includegraphics[width=0.3\textwidth,height=5cm]{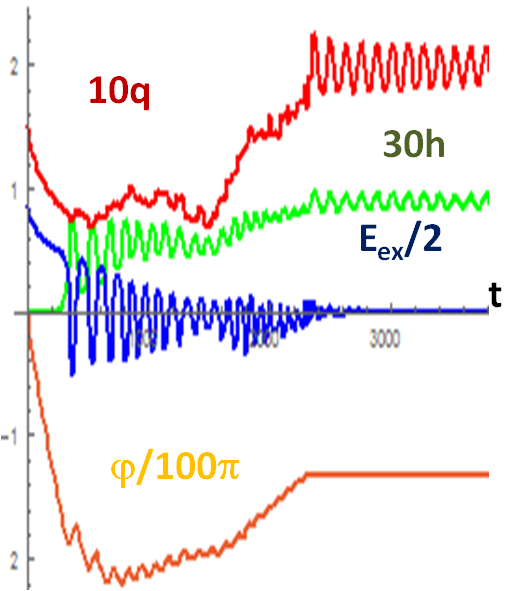}
\caption{Plots for $t$ dependencies at the subcritical pumping (left panel) and the supercritical one (right panel). $E_{ex}$ is in blue, $q$ is in red, $h$ is in green.}
\label{fig6}
\end{figure}

The Fig.6a shows results for the subcritical regime when the system is pumped from the neutral state to $q_i=0.01<q_h=0.03\ll q_I$, keeping the unperturbed initial $h_i=0$. (Recall, see Fig.2b, that  $q_I\approx 0.2$ and $|h_I|\approx 0.03$ in equilibrium of the ionic phase.) After some waiting time $t_h\approx 1000$, pronounced oscillations in $h(t)$ emerge and then decrease with a tendency to saturation at $h\approx 0.002$. But at $t^*\approx 9000$, the dynamical regime switches abruptly to weaker oscillations around $h=0$. All the way, $q(t)$ decreases monotonically while $E_{ex}$ stays nearly constant close to the unperturbed value $E_{ex}^0=0.6$. On both sides of the transition, oscillations in $E_{ex}(t)$ are small but change in character (as it could be seen by comparison with the modeling at $S=0$, which is not shown). At $t<t^*$, $E_{ex}(t)$ oscillates together with $h(t)$ at the lattice frequency $\omega$, while at $t>t^*$, the oscillations of $E_{ex}(t)$ are seen only if $S\ne 0$, and hence they are related to the macroscopic quantum interference: excitons pair creation from the vacuum admixes the basic state at $E_{ex}^0$ with states at $3E_{ex}^0$ and $-E_{ex}^0$

The Figure 6b shows the results for the supercritical regime after pumping to the level much higher than the threshold $q_h$ but still below the new equilibrium: $q_i=0.15<q_I=0.2$ (again starting with the unperturbed $h_i=0$). At short times, we see a smooth decreases in $q(t)$  and $E_{ex}(t)$ until the initiation of $h(t)$ at $t_h\approx 300$. It is followed by formation of lattice oscillations, now around a finite value of $h$ together with a finite value of the weaker oscillating $q(t)$. In this regime of an accidentally found intermediate equilibrium, the exciton energy strongly oscillates around zero such that the mean phase does not change much. After another crossover at $t^*\approx 1700$, the oscillations of $E_{ex}(t)$ make preferable excursions  to negative values: we see this from the phase, which starts anomalously growing in average.
In this regime, the excitons are preferably generated from the vacuum, and hence $q(t)$ starts to increase (with time it rises above the initial pumping level). Then, after the clearly seen lock--in transition at $t_{lock}\approx 2600$, the final equilibrium becomes apparent but with the new rise of strong oscillations in $q(t)$ and $h(t)$ provoked by this final dynamical transition.

\subsubsection{Spontaneous domain structure.}

For an extended system,  spatially homogeneous solutions may not be stable because of the interaction between the excitons and the order parameter. For a low concentration and above the BEC transition, that would be effects of self--trapping of individual excitons \cite{Rashba}. For the macroscopic description of the BEC, the effect resembles  self--focusing in nonlinear optics. In the last case, the optical "bright solitons" appear because of the negative nonlinearity --- the term $a|\Psi|^2\Psi$ with $a<0$ in the NLSE. When the NLSE for classical waves becomes the Ginzburg--Pitaevskii    equation for an ensemble of quantum Bose particles like the excitons, the negative $a$ means attraction of particles, which gives rise to the microscopic instability \cite{Nozieres:82} of the Bose gas with respect to the collapse to a liquid state (for cold atoms) or with a probable dissociation to e--h droplets for excitons (this is one more opportunity to recognize another very well--known invention by Keldysh, see \cite{Keldvsh:88}). In our case, the excitons themselves are repulsive, $a>0$; but if the order parameter is excluded (which is not possible explicitly), then the direct repulsion is overcome by  effective attraction. Microscopically, this attraction is indirect and retarded, and hence it may not lead to an instability towards the dense phase. But at larger time and space scales, the effective attraction can win, leading to spatially modulated structures.
The effect has been modeled in detail \cite{SB+NK:2014,Yi+NK+SB:2015} for a system where the excitons and the order parameter are essentially different entities, like intramolecular excitons with respect to  charge transfer in materials with the neutral--ionic  transition. The effect is also present here within the model of the excitonic insulator for the neutral--ionic  transition with pumping to charge--transfer excitons.

Below we present results of  modeling the spacio--temporal behavior  with the anomalous interaction neglected, $S=0$. In all cases, the initial wave function is taken as the lowest state in the box of the width $2L=200$ (in units of the intermolecular spacing):
$\Psi(x,0)=\Psi_i(x)=\sqrt{q_i}\cos(\pi x/2L)$. The initial maximal intensity is chosen as $q_i\approx 0.15\div 0.17$, which is well above the barrier but still below the equilibrium value $q_0=0.2$ of the ionic state, and a further $1/\sqrt{2}$ below the mean equilibrium value for the whole sample. The dimerization field $h(x,t$) is seeded as a very small value $h(x,0)=h_i(x)\sim 10^{-8}$.

Figure 7 shows the results for the homogeneous seeding $h_i(x)=const$. The density $q$ first decreases maintaining the broad initial shape. Then the self--trapping progresses, culminating at $t\approx 10000$ by a sharp rise of the central peak; soon, a rectangular shape is formed separating the system into a narrow domain of the perfect ionic phase surrounded by the wings of the nearly perfect neutral phase. The dimerization $h(x,t)$ becomes pronounced by $t\approx 300$; by $t\approx 500$, the two strong side peaks are visible in the cross section. The three-dimensional plot shows that these are passing ways emitted by the fast nucleation of  self--trapping. The later evolution resembles the one for $q(x,t)$. The difference is that oscillations of $q$ are concentrated within the narrow nucleation region.

\begin{figure}[htbp]
    \includegraphics[width=7cm]{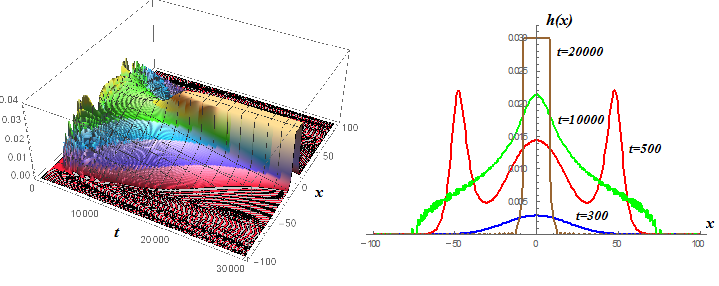} \quad
    \includegraphics[width=7cm]{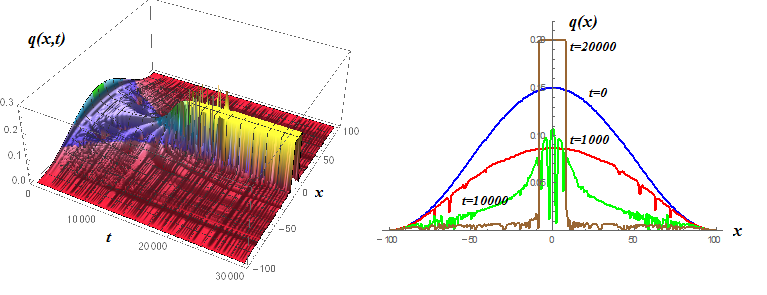}
    \caption{Plots for post--pumping evolution of $h(x,t)$ (left panel) and $q(x,t)$ (right panel) for the homogeneous seeding of $h(x,0)$.}
    \label{fig7}
\end{figure}

Because of the two--fold degeneracy with respect to the field $h$, the self--trapping direction of $h$ depends on the sign of the initial seeding of $h_i$,irrespective of how  small its magnitude is. With the always present initial inhomogeneities, the nucleation of different domains becomes inevitable. For a transparent illustration, we make the stepwise seeding $h_i(x)$ with opposite signs at two halves of the sample. The results are shown in Fig.8. \\
\\
\\
\begin{figure}[htbp]
\includegraphics[width=7cm]{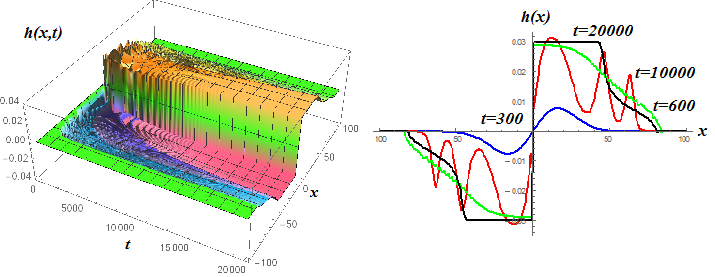} \quad
\includegraphics[width=7cm]{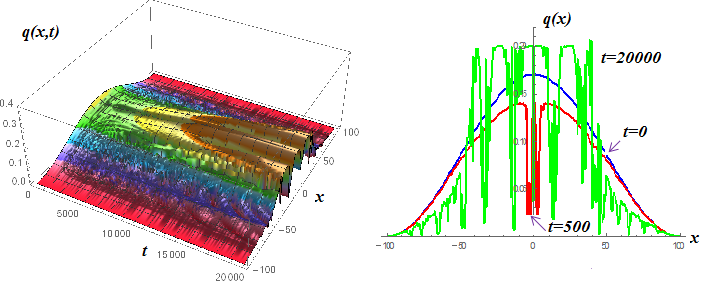}\\
\caption{Plots for post--pumping evolution of $h(x,t)$ (left panel) and $q(x,t)$ (right panel) for development of the two--domains configuration after the stepwise seeding of $h(x,0)$.}
\label{fig8}
\end{figure}

 The anti--symmetric shape of $h(x,t)$ is preserved at all $t$, and hence the domain wall (the kink--soliton) is always present around $x=0$. The humps in $h(x)$ rise to the oder of magnitude of the final scale $\pm h_0$ at $t\approx 300$. Soon, at $t\approx 600$, $h(x)$ spreads wider in the course of high--amplitude oscillations, which actually are of  dynamical origin, from a sequence of waves emitted at the early time of fast growth at small $x$. At higher $t$, the shape becomes smooth again. The final profile shows the ideally flat plateaus at $|x|<50$ of the pure ionic phase with exactly $h(x)\equiv \pm h_0=\pm 0.03$. They are surrounded by crossover layers of decreasing $h(x)$, spreading over $50<|x|<80$, which makes a difference with respect to the previous case. Beyond the sharp boundary towards the outer regions at $80<|x|$, the system stays at the pre--pumping neutral state with $h(x)\equiv 0$. The plots of $q(x,t)$ confirm the formation of the ionic phase at $|x|<50$ but the plateau is fragmented by a sequence of narrow deep rims. In contrast to $h(x,t)$, there is no sharp second boundary towards the neutral phase: $q(x)$ keeps tails comparable with the initial distribution just after the pumping.

Formation of flat sharply bounded plateaus corresponding to an ideal phase separation was not expected from a common experience with NLSE and Ginzburg--Pitaevskii    equations, where the self--focusing profiles show smooth bell--shaped humps. The difference seems to come from the threshold formation of $h$ with increasing $q$, giving rise to the kink in the dependence $V(q)$ as shown in  Fig.2.

There is an apparent correspondence between the shown pictures and the earlier schemes invoking solitons \cite{Mitani:84,Nagaosa:86}. Thus, the final domain wall between domains with opposite signs of $h$ corresponds to the always allowed, presumably spin--carrying, solitons  in \cite{Mitani:84}. The humps within the domains of a given sign of $h$, appearing here because of self--trapping, correspond to pairs of charged spinless solitons that are the walls framing the ionic string within the neutral domain \cite{Nagaosa:86}. Indeed, these solitons are always confined in pairs except  exactly at the transition temperature.  Solitons in neutral--ionic  systems in relation to the ferroelectricity will be reviewed in  more detail in\cite{SB+NK:2016}.

\section{The work to be done: a two--fluid kinetics}
The excitons forming the excitonic insulator ground state are in the condensed form by definition. In the modeling  presented above, we have additionally supposed that all the pumped excitons also form the BEC, and hence the whole ensemble can be described by a single wave function $\Psi$ of the collective state. Certainly, there is also the normal component because  the temperature is not very low and because of the initial incoherent background coming from excitons that were excited non--resonantly, with the phonon--assisted absorption of photons. Within this article, we do not address the very important question of kinetics of non condensed particles, relying on experimental facts of  fast initial equilibration. A future microscopical study can be advanced thanks to the progress in the theory of equilibration in a gas of excitons \cite{Tikhodeev:90}  and polaritons \cite{Wouters:2007,Szymanska:2006,Gardiner:2002}, and to the general understanding of a non equilibrium Bose gas motivated by problems in cold atoms \cite{Kagan:93,Kagan:94,Kagan:97}. 

There are two levels of difficulties on this way. One,  having been surpassed reasonably in the existing literature, is the kinetics of occupation numbers leading to the growing of the low--momentum peak while approaching the BEC. Another is the establishing of  coherence, allowing  the Ginzburg--Pitaevskii    description to be introduced; this final step has not been passed yet, to our knowledge, to treatable implementations, except a heavy duty numerical work \cite{Berloff:2002,Semikoz:97}. For the ideal model of the weakly interacting Bose gas, there is a fair overlap between the regime of the microscopical kinetic and the collective NLSE--based descriptions. But the price is that turbulent mixing must be taken into account in the NLSE \cite{Semikoz:97,Berloff:2002} or equivalently the Ginzburg--Pitaevskii    equation must be considered stochastically rather than  deterministically \cite{Gardiner:2002}. For applications in solid state physics,  the universality of the NLSE is not much helpful as regards the BEC of excitons and polaritons, because other channels of the relaxation become more important than collisions of bosons: emission of phonons  \cite{Tikhodeev:90} or disorder \cite{Finkelstein:13}. New features appear such as the final threshold for pumping to reach the BEC of excitons even at $T=0$ \cite{Tikhodeev:90}.

For the stationary BEC of polaritons, the theory is usually simplified by considering the normal component  as a separate quasi--equilibrium reservoir, which can be characterized by the density $n$ or the chemical potential $\mu_n$, (see \cite{Berloff:2013,Wouters:2007,Szymanska:2006,Gardiner:2002}).
While this approach, mainly reduced to the cases of the stationary pumping, will doubtfully be extended to our systems, we  briefly outline its possible application below as an absolutely minimalistic description.

A certain ground for the separation into two distinct, particle--exchanging reservoirs comes from suggesting a bottleneck --- a minimum $E_{min}$ of the kinetic energy --- where the pumped excitations accumulate after the initial rapid cooling.  It is tempting to associate $E_{min}$ with the energy of the lowest lattice mode interacting with excitons. In our case, a good candidate is the soft mode in the dip of the Kohn anomaly, which should exist as a precursor for the lattice dimerization instability, (see \cite{DAvino} and the references therein). That can  also be the Debye frequency of the acoustical spectrum; both candidates converge to $E_{min}\sim 100 \ K$.

We can make a simplifying, and quite plausible, suggestion that all reservoirs of excitons contribute additively to the order parameter: the charge transfer becomes $\Delta\rho=q+n$, where still $q=|\Psi|^{2}$. Then the system energy and the particle potential are simply  $W(q+n)$ and $V(q+n)$. Now  Gross--Pitaevskii equation (\ref{psi}) is further generalized to
\[
i\hbar\partial_{t}\Psi=
-\frac{\hbar^{2}}{2M}\partial_{x}^{2}\Psi+V(|\Psi|^{2}+n)\Psi
+(i/2)Rn\Psi-S/2\Psi^{\ast}
\]
where $R$ is a conversion rate. This has to be complemented by an equation for $n$, which we choose as a simple rate equation, 
(cf. \cite{Wouters:2007})

\[
\partial_{t}n-\partial_{x}b\partial_{x}\mu_n n=I-Rn|\Psi|^{2}~,~|\Psi|^{2}=q
\]
where $\mu_n(n)$ and $b(n)$ are the chemical potential and the mobility (see \cite{Finkelstein:13}) of normal particles. $I(t)$ is the pump intensity profile; being short, it can be omitted in favor of the initial condition $n(0)=n_{0}=\int I(t)dt$.

Since we are now considering   relatively short times, we  omit the decay terms proportional to $G$ for the total number of excitons but instead  introduce the conversion rate $R$ regulating the exchange between the reservoirs. The function $R$ must change  sign as a function of the discrepancy $\delta\mu=\mu_{n}-\mu_{c}$ of chemical potentials in the normal and  condensed subsystems. We shall adopt the simplest linear form valid at $|\delta\mu|\ll T$; otherwise, it can be generalized to $R\propto\sinh(\delta\mu/T)$ or to a more complicated nonsymmetric form. With a common definition for the chemical potential $\mu_{c}$ of the BEC, we have
\[
\mu_{c}=-\hbar\partial_{t}\varphi+
\frac{\hbar^{2}}{2M}(\partial_{x}\varphi)^{2}
~,~\mu_{n}=V(|\Psi|^{2}+n)+E_{min}~,~R=k(\mu_{n}-\mu_{c})/\hbar ,
\]
The equation of state $\mu_n(n)$ can be estimated from the standard theory of a weakly interacting Bose gas. Now, the space--independent Eqs. (3) and (4) are generalized as
\begin{eqnarray}
\hbar\dot{\varphi}=-V(q+n)+S\cos(2\varphi) ,
\\
\dot{q}=qR+Sq\sin(2\varphi)=
kqn(\dot{\varphi}+\mu_{n}/\hbar)+Sq\sin(2\varphi) ,
\nonumber \\
=(k/\hbar)qn(E_{min}+S\cos(2\varphi))+Sq\sin(2\varphi)\\
\dot{n}=P-kqn(\dot{\varphi}+\mu_{n}/\hbar)=
P-(k/\hbar)qn( E_{min}+S\cos(2\varphi)) .
\end{eqnarray}

The equation for $h$ is generalized as

\begin{equation}
K(\partial_{t}^{2}+\gamma\partial_{t})h-Ks^{2}\partial_{x}^{2}h
+d(q_{h}-q-n)h+fh^{3}=0
\end{equation}
The numerical modeling of the resulting equations and, hopefully, of more complicated dynamical--kinetic system, will be discussed elsewhere.

\section{Discussion and conclusions}

We have presented results of a phenomenological modeling for a system prone to a weakly first--order phase transition after it is exposed to the optical pumping to a high concentration of excitons. We focused on the cases where the excitation density and a thermodynamic variable present the same entity. The best--known example is the neutral--ionic  transition in donor--acceptor compounds where the charge--transfer excitons play the role of optical excitations and give the intermolecular electronic transfer as the phase transition order parameter. Both thermodynamic and dynamical effects can be described on the same root by viewing the ordered state as an excitonic insulator. Our main assumption was that a quasi--condensate of optically pumped excitons  appears sufficiently early as a macroscopic quantum state. It evolves by interacting with other degrees of freedom prone to instability,  leading to self--trapping of excitons akin to self--focusing in optics. A distinguished feature is the appearance of oscillations coming from the macroscopic quantum coherence.

Our studies are only most natural first steps in the complicated problem, and it is necessary to  quote what has not or could not be done. We have been working within a phenomenological approach that can be characterized as the one that would be valid if it could be derived microscopically.   Even within these reservations, it is desirable to also take  the normal, non condensed density of excitons and its (re)conversion (from)to the condensate into account. Indeed, temperatures in the experiments are comparable with the estimated degeneracy temperature of the BEC, and they are further enhanced in the early stage after the pumping pulse. The not--quite--resonance pumping  also contributes to the initial incoherent density. With a progressive dilution of the excitons' density,  the BEC transition must be passed back even at low temperature.

Our primary emphasis was upon the quantum nature of the exciton motion, which forces their delocalization into plain waves. That would happen inevitably for an ideal resonance pumping when a single photon creates a single exciton with the momentum $k=0$. In reality, a large part of photons  is absorbed with an access energy, which gives rise to the exciton in a complex with other modes whose total momentum is still zero, but the exciton acquires a momentum and the associated kinetic energy. The initial relaxation by collisions leaves the exciton as a wave packet rather than a pure state, which still cannot be viewed as localized at a single molecule or a dimer, as it is commonly pictured  in the scenario of "falling dominos".  The loss of the kinetic energy from such a sharp localization will be $>0.1 eV$  as estimated from the exciton bandwidth in  optical absorption. Then, with cooling below $\sim$1000K, the exciton  descends to the plane wave state at the bottom of its band. The smooth localization, which we have modeled here and previously \cite{Brazovskii:2014-JSNM}, then develops as self--trapping; its length is determined by the balance between gaining the potential energy and loosing the quantum kinetic energy of the exciton. 
In either case, the kinetics of cooling must be taken into account and the incoherent component of the exciton ensemble should be added.

Microscopic theories of the dynamical BEC offer an important experience to learn, being motivated by problems in polaritons (see, e.g., \cite{Wouters:2007,Szymanska:2006} and more references  in  review \cite{Berloff:2013}) and cold atoms \cite{Gardiner:2002,Kagan:93,Kagan:97}.
The theory is able to reproduce the growing of occupation numbers at lowest energies, as it was beautifully traced in experiments with cold atoms. But establishing  phase coherence has not been clearly derived yet;  the  Ginzburg--Pitaevskii--type equations appear from the microscopic theory in a stochastic rather than  deterministic form.

Contrary to a simple and treatable microscopic nature of excitons in semiconductors, polaritons, and cold atoms, we here  face a very complicated origin of the exciton and of the affected instabilities: intramolecular electronic correlations as a source of charge transfer against the Coulomb and kinetic energies, cooperative correlations leading to the spin--Peierls instability of  lattice dimerization.

Among applications to neutral--ionic transitions in organic donor--acceptor materials, we face the fermionization of excitons as initially repulsive Bose particles, because of the rather one-dimensional structure of these materials. More specifically, there is an unresolved problem of the commonly accepted interpretation of the absorption peak at $0.6 eV$ as the charge--transfer exciton     energy $E_{ex}$, because it keeps nearly the same position in the ionic phase as in the neutral one. This is not compatible with our, and probably any, phenomenological theory: the strong effect of excitons on the equilibrium value of $\rho$ implies the reciprocal effect of $\rho$ on $E_{ex}$. The extensive microscopic modeling in \cite{Yonemitsu:2012} also shows the expected strong dependence of $E_{ex}$ on $\rho$.

We believe that the suggested picture, the approach, and the illustrations will encourage a more solid theoretical work and will stimulate experimental studies of PIPT in systems possessing features of the excitonic insulator and/or allowing  pumping to the excitation modes coupled to a parameter of a nearby phase transition.

\begin{acknowledgments}
The authors are grateful to Prof. H. Okamoto for introducing them to the field of optically induced neutral--ionic transitions and for numerous discussions. One of the authors (S.B.) wishes to acknowledge funding from the ERC AdG Trajectory.
\end{acknowledgments}

\newpage
{\bf{Appendix}
\\
{Physical parameters and estimations.}}

We must relate the constants in eqs. (\ref{psi}), (\ref{V(q,h)}), and (\ref{h}) with physical parameters and estimate their values.

$q_{I}$: the charge transfer goes from $\rho_{N}=0.32$ to $\rho_{I}=0.52$,
whence $q_{I}=0.2$.

$E_{ex}^0$: the charge--transfer exciton energy $E_{ex}^0=0.6$ eV is known.

$a$: $a$ is the parameter of exciton interaction, which is bounded from above by the energy of  exciton dissociation, that is, $a\sim10^{-1}$ eV.

$c$ and $q_{h}$: at the transition temperature $T_{NI}$, $W_{I}=W_{N}$ and $dW_{I}/dq=0$ which yields $q_{h}\approx0.01$ and $c\approx1.4$.

$f$: in units of $d=7.4\AA$, the dimerization in the ionic phase is $h_{I}=0.03$ \cite{Kobayashi}. Knowing $q_{I}$ and $h_{I}$, we obtain $f/c=(q_{I}/q_{h}-1)/h_{I}^{2}$.

$b$: the requirement for the correct boundary between neutral and ionic states gives  
$b\approx 50$. We note the much higher value of $b$ compared with the estimation 
$b\propto a/q_{I}=5a$ from considering the energy term $bq^{3}$ as an unharmonism with respect to $aq^{2}$.

$m$: the experimental width of the exciton absorption line gives the estimate $\hbar^{2}/2md^{2}=0.2-0.3$ eV.

$\omega$, $\gamma$: the period of coherent oscillations is $T_{osc}=0.6ps$, and the dimer mode frequency is then $\omega=2\pi/T_{osc}\approx10^{-2} fs^{-1}$. Their relaxation time is  $\tau_{h}=(3-7)$ ps \cite{Okamoto}; taking it as $5$ ps, the damping parameter is $\gamma=1/\omega\tau_{h}\approx 0.02$.

$G$: the inverse lifetime of residual ionic segments  is $\tau_{I}=20$ ps \cite{Okamoto}, and therefore  the modeling should yield the dynamical phase transition at a time of the order of $10T_{osc}$. This requires  $G\sim 10^{-3}$.

$A$: it is expressed via $\omega$ and the sound velocity $s$ as $A/c=s^{2}/\omega^{2}\sim0.02$ from the estimate $s\propto d/T_{osc}\approx10^{5}$ cm/s, which is on the scale of values measured for other charge--transfer     compounds.

\end{document}